\documentclass[nofootinbib,aps,10pt,twocolumn]{revtex4-1}

\usepackage{array,amsmath,amsthm}
\usepackage{mathtools}
\usepackage{graphicx}
\usepackage{color}
\usepackage{amssymb}

\begin{document}
\title{Electroweak Absolute, Meta-, and Thermal Stability in Neutrino Mass Models}
\author{Manfred Lindner}
\email{lindner@mpi-hd.mpg.de}
\affiliation{Particle and Astro-Particle Physics Division \\
Max-Planck Institut fuer Kernphysik {\rm{(MPIK)}} \\
Saupfercheckweg 1, 69117 Heidelberg, Germany}
\author{Hiren H. Patel}
\email{hiren.patel@mpi-hd.mpg.de}
\affiliation{Particle and Astro-Particle Physics Division \\
Max-Planck Institut fuer Kernphysik {\rm{(MPIK)}} \\
Saupfercheckweg 1, 69117 Heidelberg, Germany}
\author{Branimir Radov\v{c}i\'{c}}
\email{radovcic@mpi-hd.mpg.de}
\affiliation{Particle and Astro-Particle Physics Division \\
Max-Planck Institut fuer Kernphysik {\rm{(MPIK)}} \\
Saupfercheckweg 1, 69117 Heidelberg, Germany}
\begin{abstract}
We analyze the stability of the electroweak vacuum in neutrino mass models containing right handed neutrinos or fermionic isotriplets.  In addition to considering absolute stability, we place limits on the Yukawa couplings of new fermions based on metastability and thermal stability in the early Universe.  Our results reveal that the upper limits on the neutrino Yukawa couplings can change significantly when the top quark mass is allowed to vary within the experimental range of uncertainty in its determination.
\end{abstract}
\maketitle
\section{Introduction}
With the recent discovery \cite{Aad:2012tfa,Chatrchyan:2012xdj} of the Higgs boson at a mass of 125 GeV, it is possible to answer the question of vacuum stability \cite{Lindner:1988ww} in the standard model up to any given scale of new physics.  The exceptionally large top quark coupling to the Higgs field renders the electroweak vacuum  metastable, but with a lifetime larger than the age of the universe \cite{Buttazzo:2013uya}.  At extreme temperatures present in the early universe, thermal fluctuations can cause transitions away from the electroweak vacuum \cite{Anderson:1990aa, Arnold:1991cv, Espinosa:1995se, Espinosa:2007qp}, and puts an upper limit on the reheating temperature at the end of inflation assuming no new physics beyond the standard model \cite{Rose:2015lna}.

It is possible that the presence of new unknown fermions sizably coupled to the Higgs can exacerbate the already precarious state of the electroweak vacuum.  Models of neutrino mass generation typically contain additional fermionic degrees of freedom coupled to the standard model Higgs.  In this paper, we explore the effects of these new fermions on electroweak vacuum stability, metastability, and stability against thermal fluctuations in the early Universe.

In the past, there has been some work in this area.  In \cite{Casas:1999cd}, the absolute stability bounds in high scale type-I seesaw models were considered. The analysis was extended in \cite{EliasMiro:2011aa} to include metastability and thermal stability bounds, and in \cite{Rodejohann:2012px, Bambhaniya:2014hla} to cover the possibility of large neutrino Yukawa couplings due to non-trivial flavor structure.  Most recently, constraints from vacuum metastability were studied in the context of inverse seesaw model \cite{Rose:2015fua} and linear seesaw models \cite{Khan:2012zw}.  In \cite{Gogoladze:2008ak,He:2012ub}, constraints on type-III seesaw models were found by requiring absolute stability.

We separately analyze the impact of right handed neutrinos (iso\-sing\-lets) or fermion iso\-trip\-lets on the absolute stability of the electroweak vacuum.  A novel feature of our analysis is that it accommodates the possibility that these new degrees of freedom are at the low scale, and in a manner that is general with respect to the specific flavor structure of the Yukawa coupling matrix.  Furthermore, this allows us to clearly present results as a function of the top quark pole mass.  We include vacuum metastability and thermal stability analyses for both cases.  And, while our findings confirm literature for the case of high scale right handed neutrinos, our results for fermion isotriplets and low scale right handed neutrinos are new.

In section II we briefly describe the theory behind our analysis of vacuum stability.  In section III and IV we analyze and derive vacuum stability bounds in models with right handed neutrinos and in models with isospin triplets.  We discuss the implications of our findings on experimental observables in Section V, and we conclude in section VI.

\section{Theory}
In this section, we summarize the theory underlying the analysis of vacuum stability at zero and finite temperature.  For further details, the interested reader should consult the papers referenced in this section.

\subsection{Absolute vacuum stability and metastability}
We analyze the vacuum structure and the question of metastability by closely following the methods of \cite{Isidori:2001bm, Buttazzo:2013uya} applied to the case of the standard model. 

In our analysis, we assume that there is no new physics that significantly modifies the form of the Higgs potential apart from the additional degrees of freedom needed to generate neutrino masses below the Planck scale.  Absolute electroweak vacuum stability is lost if the value of the Higgs field effective potential
\begin{equation}\label{eq:SMtree}
V(h) = \frac{1}{4} \lambda(\mu) h^4
\end{equation}
dips below that of the electroweak vacuum.  This requirement is well-approximated by the condition that the effective coupling $\lambda(\mu)$ remain positive for the scale $\mu$ up to the Planck scale.  The Higgs quadratic coupling can be safely neglected because, for values of the electroweak parameters near their measured values,  destabilization occurs at values of $\mu$ much higher than the electroweak scale. We use the two-loop standard model beta functions, and the two-loop relationship between  the $\overline{\text{MS}}$ and pole masses for the Higgs boson and the top quark \cite{Buttazzo:2013uya}.

Although the electroweak vacuum may not be the global minimum of the effective potential, it may be a long-lived metastable state with a lifetime much larger than the age of the universe.  The tunneling rate is computed in the semiclassical approximation based on potential in (\ref{eq:SMtree}).  The bounce solution and action associated with tunneling are \cite{Isidori:2001bm}
\begin{equation}
h_\text{B}(r) = \sqrt{\frac{2}{|\lambda(\mu)|}} \frac{2R}{r^2 + R^2},\qquad
S_\text{B}(h_\text{B}) = \frac{8\pi^2}{3|\lambda(\mu)|}\,.
\end{equation}
Here, $R$ characterizes the size of the ``bounce'' solution, and is undetermined on account of the scale invariance exhibited by the potential in (\ref{eq:SMtree}) at the classical level.  However, the anomalous scaling of $\lambda$ under the renormalization group breaks the scale invariance.  To minimize the impact of large logarithms at higher order, we set the renormalization scale $\mu= R^{-1}$.  Then the probability for tunneling is given by
\begin{equation}
p \approx \text{max}_R \frac{\text{vol.}}{R^4} \exp \Big[-\frac{8\pi^2}{3|\lambda(R^{-1})|}\Big]\,,
\end{equation}
where $\text{``vol.''} = 0.15 H_0^{-4}$ is volume of our past lightcone.  The fraction of space in the true vacuum goes like $e^{-p}$ \cite{Guth:1981uk}, and our criteria for stability is $p \gtrsim 1$.

\subsection{Thermal stability}
We are also interested in the question of stability against thermal fluctuations in the early Universe.  We follow the procedure laid out in \cite{Arnold:1991cv} which is an application of Kramer's barrier crossing formalism \cite{Kramers:1940zz} to the standard model Higgs potential.  It is sufficient for us to use the high-temperature effective action retaining just the $\mathcal{O}(T^2)$ terms in the effective potential,
\begin{equation}
V_\text{eff}(h,T) = \frac{1}{4}\lambda(\mu) h^4 + \frac{1}{2}\kappa^2(\mu) h^2 T^2,
\end{equation}
where in the standard model,
\begin{equation}\label{eq:thermCoeffSM}
\kappa^2 = \frac{1}{16}(g'^2 + 3g^2 + 4y_t^2 + 8\lambda)\,.
\end{equation}
Although the impact of dropping subleading terms in the high-$T$ expansion leads to a substantial underestimate of the crossing probability, its impact on the bounds of the model parameters is miniscule \cite{Arnold:1991cv,Rose:2015lna}.  Furthermore, because the subleading terms are gauge dependent, this approximation has the additional advantage of maintaining gauge independence in our final results \cite{Patel:2011th}.

The critical bubble profile is obtained by minimizing the temperature dependent energy functional
\begin{equation}\textstyle
E = \frac{\kappa T}{|\lambda|} \int d^3 \bar{r} \big[\frac{1}{2}(\nabla \bar h)^2 + \frac{1}{2}{\bar h}^2-\frac{1}{4}{\bar h}^4\big]\,,
\end{equation}
subject to the boundary condition
\begin{equation}\textstyle
\frac{d\bar h (\bar r)}{d\bar r}\big|_{\bar r = 0} = 0,\qquad \bar h (\bar r)\big|_{\bar r = \infty} = 0\,,\\
\end{equation}
where $\bar{r}=\kappa T |\mathbf{x}|$ and $\bar h = \sqrt{|\lambda|}h/(\kappa T)$ are the dimensionless radial coordinate and Higgs field strength.  
The bounce profile and the dimensionless integral are computed numerically, yielding the energy of the critical bubble
\begin{equation}\label{eq:bubbenergy}
E_\text{B} = 6.015\pi \frac{\kappa(\mu) T}{|\lambda(\mu)|}\,.
\end{equation}
To minimize the impact of large logarithms at higher order, we take the renormalization scale $\mu= T$.  The differential probability for thermal crossing in the radiation dominated Universe is
\begin{equation}
\frac{dp}{d T} \approx (t_0 T_0)^3 \frac{m_\text{Pl}}{T^2} e^{-E_\text{B}/T}\,,
\end{equation}
where $t_0$ is the age of the Universe and $T_0$ is the current CMB temperature.  We obtain the total probability by numerical integration starting from the highest temperature achieved in the universe $T_\text{max}$.  The integral is dominated near temperatures corresponding to scales when $|\lambda|$ is maximized, making the energy of the critical bubble in (\ref{eq:bubbenergy}) its smallest.

Having established our methods of computation, in the next two sections we analyze the impact of adding additional fermions in neutrino mass models to the stability of the electroweak vacuum at zero and finite temperature. 

\section{Right handed neutrinos}
In this section, we analyze and derive the electroweak vacuum stability bounds coming from fermionic singlets such as from right handed neutrinos.  Sizable Yukawa couplings to the right handed neutrinos appear in type-I seesaw models \cite{Minkowski:1977sc,*Yanagida:1979as,*GellMann:1980vs,*Mohapatra:1979ia} with peculiar flavor structure, or in inverse \cite{Mohapatra:1986bd,GonzalezGarcia:1988rw} or linear seesaw models \cite{Malinsky:2005bi}.

The relevant part of the Lagrangian governing the interactions of three right handed neutrinos $\nu_R$ is given by 
\begin{equation}\label{eq:singletLagrangian}
\mathcal{L} = -\overline{\ell_L}\tilde{H}{\mathbb{Y}_\nu}\nu_R - \frac{1}{2}\overline{\nu_R}\mathbb{M}_R\nu_R^c + \text{c.c.}\,,
\end{equation}
where $\mathbb{Y}_\nu$ is the neutrino Yukawa coupling matrix, and $\mathbb{M}_R$ is the right-handed neutrino mass matrix.

The coupling of the right handed neutrinos to the standard model Higgs affects the running of the quartic coupling $\lambda$ through its modification of various renormalization group equations (RGEs) at scales above $\mathbb{M}_R$. At one loop order the modifications to the beta functions for $\lambda$ and the top quark Yukawa coupling $y_t$ are \cite{Grzadkowski:1987tf,Pirogov:1998tj}
\begin{align}\label{eq:rgeLambda}
\Delta \beta_\lambda &= \frac{1}{(4\pi)^2}\Big(4\lambda \text{Tr}\big[\mathbb{Y}_\nu^\dag \mathbb{Y}_\nu\big] - 2 \text{Tr}\big[\mathbb{Y}_\nu^\dag \mathbb{Y}_\nu\mathbb{Y}_\nu^\dag \mathbb{Y}_\nu\big]\Big)\,,\\
\Delta \beta_{y_t} &= \frac{1}{(4\pi)^2} y_t \text{Tr}\big[\mathbb{Y}_\nu^\dag \mathbb{Y}_\nu\big] \,.
\end{align}
The beta function for neutrino Yukawa couplings is given by
\begin{multline*}
[\beta_{\mathbb{Y}_\nu}]_{ij} = \frac{1}{(4\pi)^2}\Big[\frac{3}{2} (\mathbb{Y}_\nu \mathbb{Y}_\nu^\dag \mathbb{Y}_\nu)_{ij} \\ + (\mathbb{Y}_\nu)_{ij}\Big(\text{Tr}\big[\mathbb{Y}_\nu^\dag \mathbb{Y}_\nu\big]
 + 3y_t^2 - \frac{3}{4}g'^2-\frac{9}{4}g^2\Big)\Big]\,,
\end{multline*}
from which it follows the beta function of the trace of the product of neutrino Yukawa coupling matrix,
\begin{multline}\label{eq:rgeTrYYsinglet}
\beta_{\text{Tr}[\mathbb{Y}^\dag_\nu \mathbb{Y}_\nu]} = \frac{1}{(4\pi)^2}\Big[3\text{Tr}[\mathbb{Y}_\nu^\dag \mathbb{Y}_\nu \mathbb{Y}_\nu^\dag \mathbb{Y}_\nu] 
+ 2\big(\text{Tr}\big[\mathbb{Y}_\nu^\dag \mathbb{Y}_\nu\big]\big)^2
 \\
 +\text{Tr}\big[\mathbb{Y}_\nu^\dag \mathbb{Y}_\nu\big]\big( 6y_t^2 - \frac{3}{2}g'^2-\frac{9}{2}g^2\big)\Big]\,.
\end{multline}
For small $\lambda$ the dominant contribution to the modified beta function $\Delta\beta_\lambda$ is the second term in (\ref{eq:rgeLambda}), which drives it downwards at high energies.

We would like to solve the RGEs without specifying all the elements of $\mathbb{Y}_\nu$ separately.  Therefore, we attempt to draw a simplifying connection between the traces $\text{Tr}[\mathbb{Y}^\dag_\nu \mathbb{Y}_\nu]$ and $\text{Tr}[\mathbb{Y}_\nu^\dag \mathbb{Y}_\nu \mathbb{Y}_\nu^\dag \mathbb{Y}_\nu]$ appearing in (\ref{eq:rgeTrYYsinglet}).  To effect this, we consider two distinct cases.

In the first case, we assume that one of the right handed neutrinos couples more strongly to the standard model Higgs, allowing us to take
\begin{equation}\label{eq:domCouplRelation}
\text{Tr}\big[(\mathbb{Y}_\nu^\dag \mathbb{Y}_\nu)\big]\rightarrow y_\nu^2\,, \qquad \text{Tr}\big[\mathbb{Y}_\nu^\dag \mathbb{Y}_\nu \mathbb{Y}_\nu^\dag \mathbb{Y}_\nu\big]\rightarrow y_\nu^4\,,
\end{equation}
where $y_\nu$ is the effective Yukawa coupling to that neutrino.  This configuration can be realized in high-scale type-I seesaw model, or in low-scale inverse, double, or linear seesaw models.

\begin{figure}
\includegraphics[width=0.8\columnwidth]{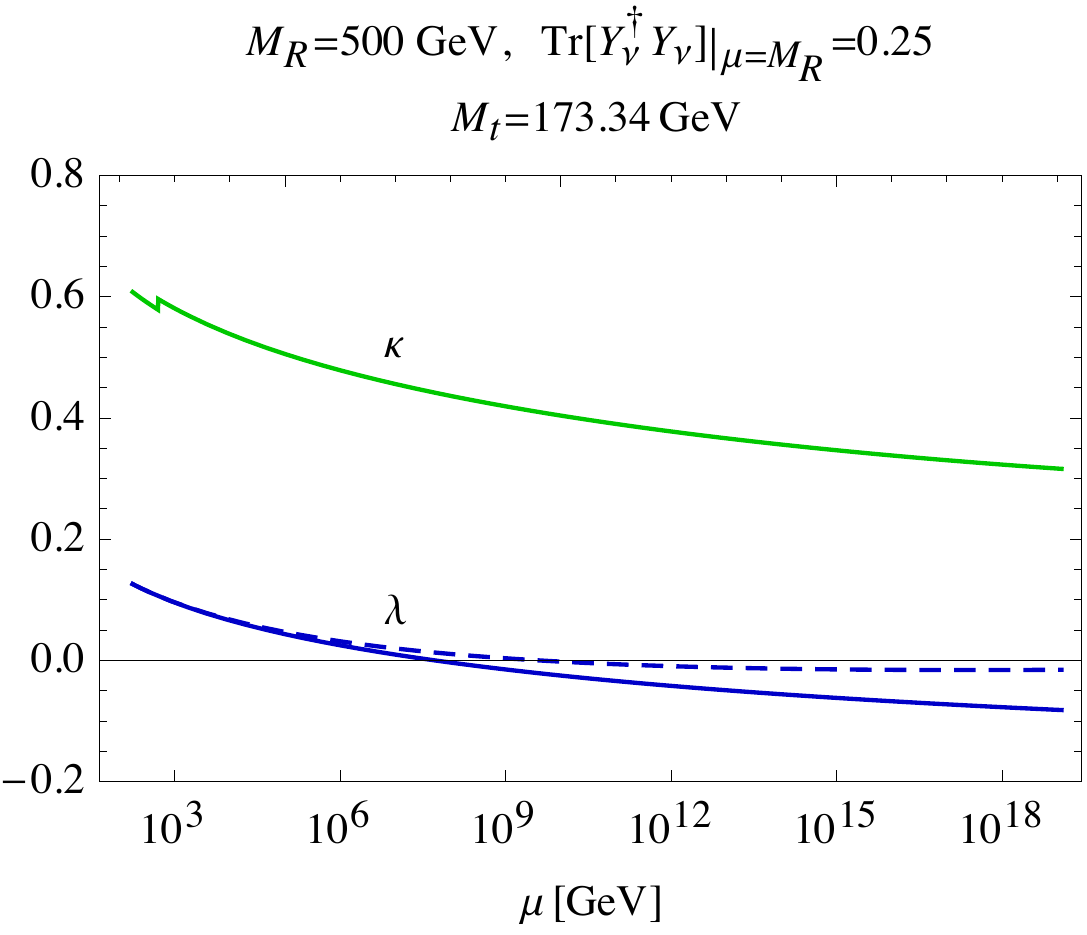}
\caption{The renormalization group evolution of the Higgs quartic coupling $\lambda$ in the presence of right handed neutrinos (solid blue) and in the standard model (dashed blue).  In green is the thermal coefficient $\kappa$ defined in (\ref{eq:thermCoeffSM}) and modified by (\ref{eq:thermalCoefficientSinglet}).}\label{fig:rgeSinglet}
\end{figure}

However, this assumption is unlikely to hold in low scale type-I seesaw unless the neutrino Yukawa couplings are so small that they have negligible impact on the Higgs potential.  This prompt us to treat this case separately, as follows.  We start with the general parametrization of the Yukawa coupling matrix given by \cite{Casas:2001sr}
\begin{equation}
\mathbb{Y}_\nu = \frac{\sqrt{2}}{v}\sqrt{\mathbb{M}_R} \mathbb{R} \sqrt{\mathbb{M}_\nu} U_\text{PMNS}^\dag\,,
\end{equation}
where $v=246$ GeV is the standard model Higgs {\sc vev}, and $\mathbb{M}_\nu = \text{diag}(m_1,m_2,m_3)$ is the diagonalized light-neutrino mass matrix.  Following \cite{Pascoli:2003rq}, we parametrize the complex orthogonal matrix $\mathbb{R}$ in terms of its polar decomposition $\mathbb{R} = \mathbb{O} \exp [i \mathbb{A}]$, where $\mathbb{O}$ is real orthogonal and $\mathbb{A}$ is real antisymmetric.  For simplicity, we assume degeneracy among the right handed neutrino masses $M_R \equiv M_1 = M_2 = M_3$.  Then, the Yukawa matrix traces can be computed as
\begin{align}
\label{eq:TrYY1}
\text{Tr}[\mathbb{Y}^\dag_\nu \mathbb{Y}_\nu] &= \frac{2M_R}{v^2}\text{Tr}\big[\mathbb{M}_\nu e^{2i\mathbb{A}}\big]\,,\\
\label{eq:TrYYYY1}
\text{Tr}\big[\mathbb{Y}_\nu^\dag \mathbb{Y}_\nu \mathbb{Y}_\nu^\dag \mathbb{Y}_\nu\big] &= \Big(\frac{2M_R}{v^2}\Big)^2\text{Tr}\big[\mathbb{M}_\nu e^{2i\mathbb{A}}\mathbb{M}_\nu e^{2i\mathbb{A}}\big]\,,
\end{align}
and are independent of $\mathbb{O}$ and $U_\text{PMNS}$.  Consider for now the case where the light neutrinos are degenerate $m_\nu \equiv m_1\approx m_2 \approx m_3$.  Recalling that the eigenvalues of an antisymmetric $3\times3$ matrix $\mathbb{A}$ can be written as $\{0,\, \pm i \rho\}$, where $\rho = \sqrt{-\frac{1}{2}\text{Tr}[\mathbb{A}\mathbb{A}]}$ is positive semidefinite, the traces in (\ref{eq:TrYY1}) and (\ref{eq:TrYYYY1}) can be written in terms of the sums of the eigenvalues of the matrix exponentials,
\begin{align}
\label{eq:TrYY2}
\text{Tr}[\mathbb{Y}^\dag_\nu \mathbb{Y}_\nu] &= \frac{2M_R}{v^2} m_\nu(e^{2\rho} + 1 + e^{-2\rho})\,,\\
\label{eq:TrYYYY2}
\text{Tr}\big[\mathbb{Y}_\nu^\dag \mathbb{Y}_\nu \mathbb{Y}_\nu^\dag \mathbb{Y}_\nu\big] &= \Big(\frac{2M_R}{v^2}\Big)^2 m_\nu^2 (e^{2\rho} + 1 + e^{-2\rho})^2\,.
\end{align}
Given that these traces are what appear in the RGE for $\lambda$, it has a sizable effect on vacuum stability only if they are large, and hence when $\rho \gg 1$.  But it is precisely in this limit that the final two terms in parenthesis $1+e^{-2\rho}$ are negligible compared to the first term $e^{+2\rho}$.  Upon dropping these terms we obtain the useful relationship
\begin{equation}\label{eq:TrYrelationship}
\text{Tr}\big[\mathbb{Y}_\nu^\dag \mathbb{Y}_\nu \mathbb{Y}_\nu^\dag \mathbb{Y}_\nu\big] \approx \Big(\text{Tr}\big[\mathbb{Y}_\nu^\dag \mathbb{Y}_\nu\big]\Big)^2\,.
\end{equation}
We find that this relationship also holds for normal and inverted mass hierarchies following similar considerations.  Therefore, we can use this approximation in (\ref{eq:rgeTrYYsinglet}) to solve the set of RGEs without knowing the detailed structure of the Yukawa matrix.

At temperatures above $M_R$, the Higgs thermal mass coefficient (\ref{eq:thermCoeffSM}) picks up an additional contribution from the right handed neutrinos given by
\begin{equation}\label{eq:thermalCoefficientSinglet}
\delta \kappa^2 = \frac{1}{12}\text{Tr}\big[\mathbb{Y}_\nu^\dag \mathbb{Y}_\nu\big]\,.
\end{equation}

In Fig. \ref{fig:rgeSinglet} we show how the renormalization group evolution of the Higgs quartic coupling $\lambda$ in this model compares with that of the standard model, for $M_R = 500$ GeV, and $\text{Tr}\big[\mathbb{Y}_\nu^\dag \mathbb{Y}_\nu\big]\big|_{\mu=M_R} = 0.25$.  On the same plot, we show how the thermal coefficient $\kappa$ runs with scale.  Notice that the presence of right handed neutrinos causes $\lambda$ to run more negatively, further destabilizing the Higgs potential.  This will lead to limits on the neutrino Yukawa couplings.

Because of the sensitivity of the running of $\lambda$ to the top quark pole mass and relatively large experimental uncertainty in its value  $M_t = 173.34 \pm 0.76$ GeV \cite{ATLAS:2014wva} and additional theoretical uncertainty, we leave the top quark mass as a free parameter when placing limits on neutrino Yukawa couplings.  We fix the Higgs mass to be $m_H = 125.09$ GeV \cite{Aad:2015zhl}.  

\begin{figure}
\includegraphics[width=0.8\columnwidth]{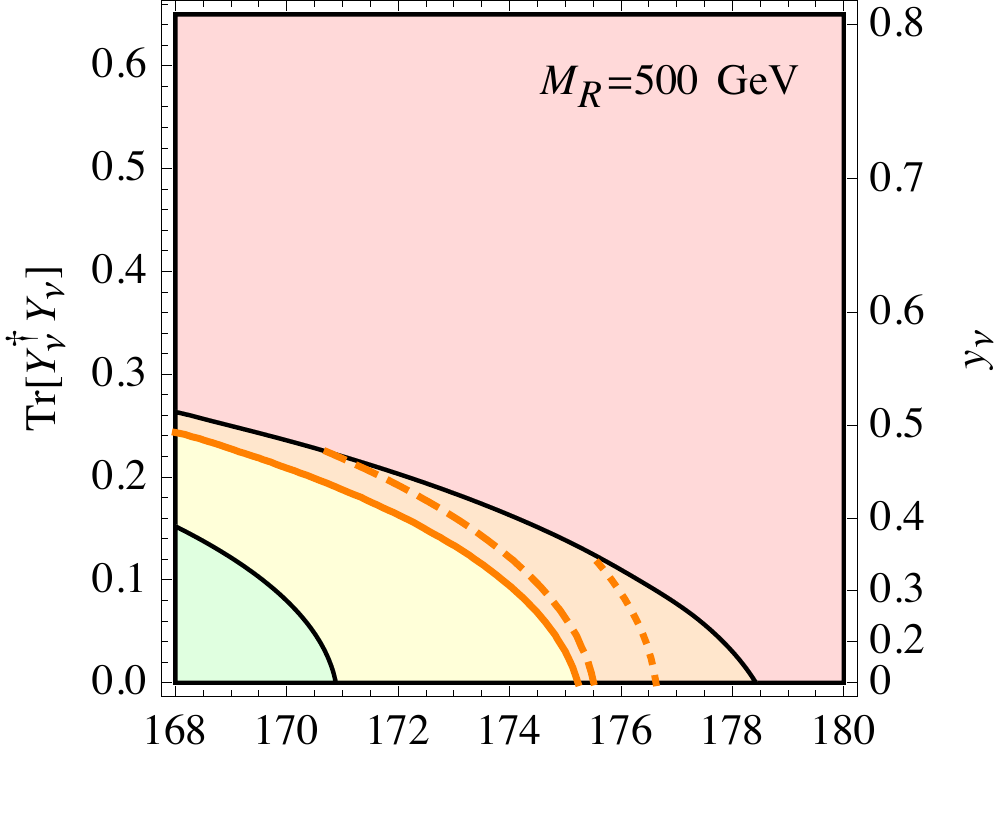}
\includegraphics[width=0.8\columnwidth]{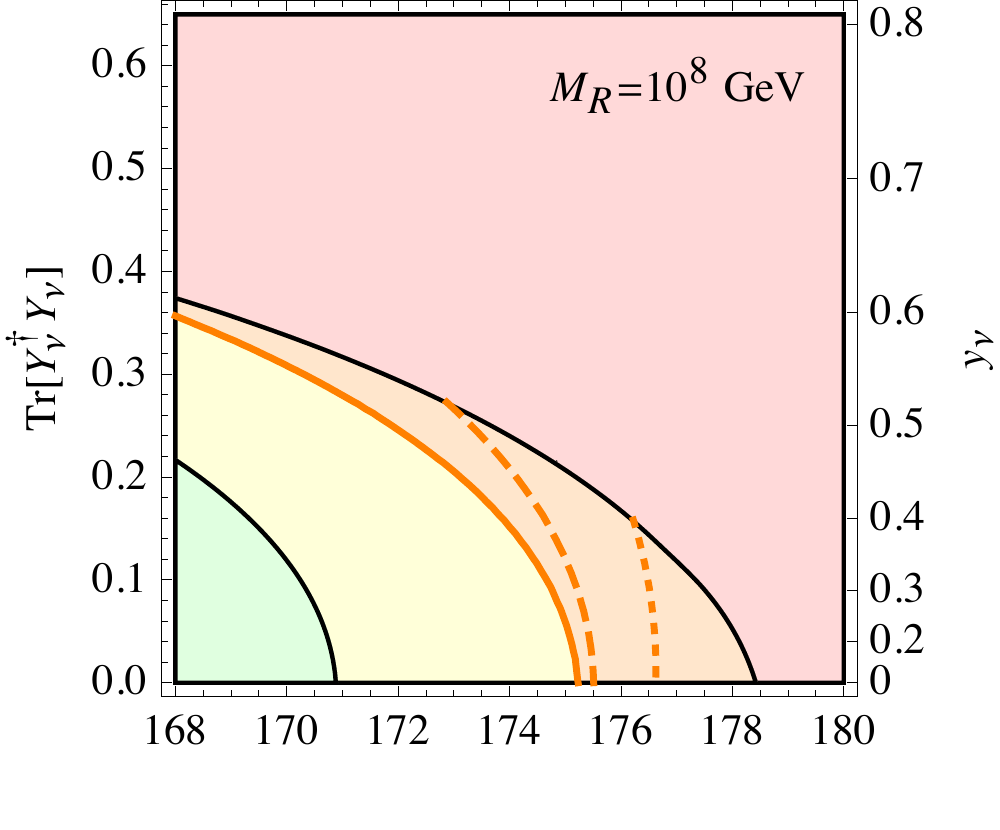}
\includegraphics[width=0.8\columnwidth]{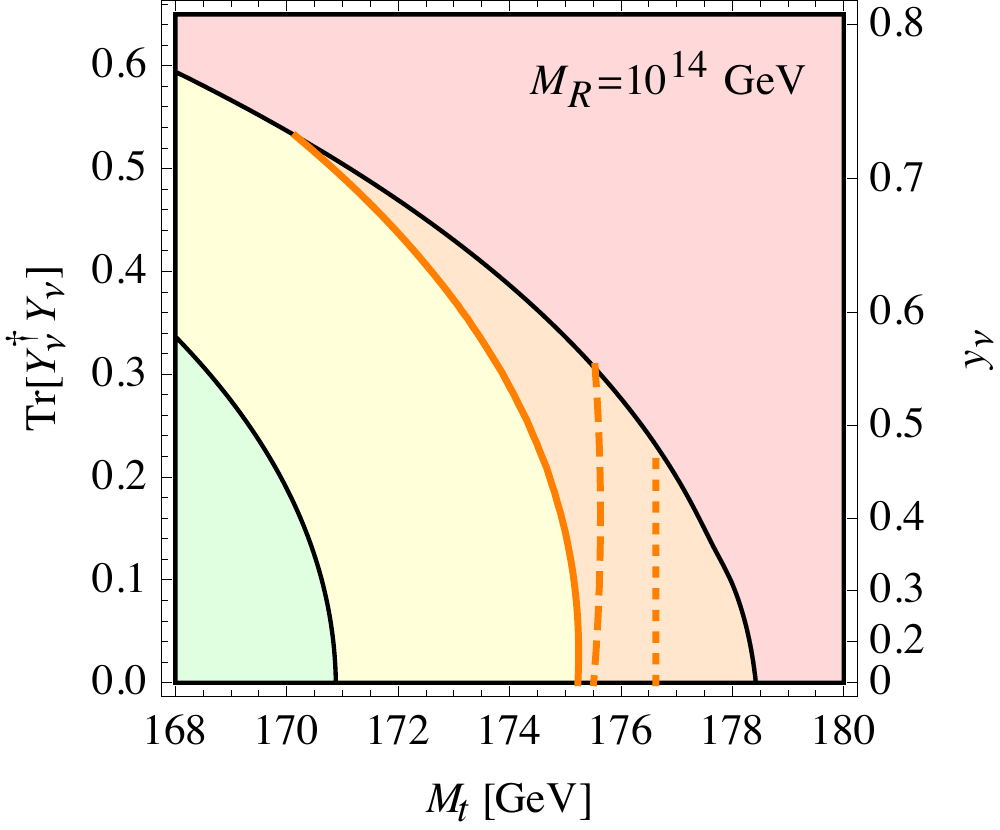}
\caption{Phase diagram of the standard model coupled to three right-handed neutrinos in the plane of $M_t$ and $y_\nu$, for representative values of $M_R$.  The plane is divided into regions of absolute stability (green), zero temperature quantum mechanical tunneling metastability (yellow and orange), and instability (red).  The region shaded in orange corresponds to an instability due to thermal transitions in the early Universe assuming a maximum temperature achieved in the Universe of $T_\text{max}=10^{18}$.  The orange region shrinks according to the dashed and dotted orange lines for lower maximum temperatures of $10^{15}$ GeV and $10^{12}$ GeV, respectively.  Note that at $2\sigma$ C.L., the top quark pole mass is $\sim 171$--$175$ GeV.}\label{fig:PDsinglet}
\end{figure}

In Fig. \ref{fig:PDsinglet}, we show the phase diagram in the \mbox{$M_t$--$y_\nu$} plane for this class of models for three choices of the right-handed neutrino mass $M_R = \{500, 10^8, 10^{14}\}$ GeV.  We show both, the combination $\text{Tr}\big[\mathbb{Y}_\nu^\dag \mathbb{Y}_\nu\big]\big|_{\mu=M_R}$ on the left axis, and $y_\nu$ on the right axis, corresponding to the two cases of the treatment of the neutrino Yukawa matrix, respectively.  The region in green is where the electroweak vacuum is absolutely stable (below the Planck scale).  In yellow and orange regions, the electroweak vacuum is metastable but with a quantum tunneling lifetime that is longer than the age of the Universe; in the red region the lifetime is shorter than the age of the Universe.  In the orange region, thermal fluctuations lead to transitions away from the electroweak vacuum in the early Universe, with the lines correspond to different maximum temperature achieved in the universe, as indicated in the caption.

Our findings confirm the following results in the literature \cite{Casas:1999cd,EliasMiro:2011aa,Rodejohann:2012px,Bambhaniya:2014hla}:

\begin{enumerate}
\item The bounds on neutrino Yukawa couplings are only logarithmically sensitive to $M_R$. 

\item The neutrino Yukawa couplings play a similar role as does that of the top quark by also destabilizing the electroweak vacuum, implying bounds on these couplings.
\end{enumerate}
However, we add the following point:
\begin{enumerate}
\setcounter{enumi}{2}
\item The sensitivity of these bounds to the precise value of the top quark mass is substantially greater.  We can readily understand the sensitivity by inspecting the behavior of $\beta_\lambda$ with respect to variations in the coupling constants $y_t$ and $y_\nu$:
\begin{equation*}
\delta \beta_\lambda \sim y_t^3 \delta y_t + y_\nu^3 \delta y_\nu
\end{equation*}
Since the value of $y_t\approx 1$ is bigger than the values of $y_\nu$ near the bounds, the coefficient in front of $\delta y_t$ is much bigger, explaining the higher sensitivity.

\end{enumerate}

\section{Case: Fermionic triplet}
In this section, we turn our attention to models containing fermionic isotriplets.  As before, sizable Yukawa couplings to the fermionic triplet appear in type-III seesaw models \cite{Foot:1988aq} with peculiar flavor structure, or in inverse seesaw models.

The Lagrangian governing the coupling of three isospin triplets $\Sigma_R$ to the standard model Higgs is
\begin{equation}\label{eq:tripletLagrangian}
\mathcal{L} = - \sqrt{2} \bar\ell_L \mathbb{Y}_\Sigma^\dag \Sigma_R \tilde H + \frac{1}{2}\text{Tr}\big[\bar\Sigma_R \mathbb{M}_\Sigma \Sigma_R\big]+\text{c.c.}
\end{equation}
For simplicity, we assume that the isotriplets are mass degenerate, $M_\Sigma \equiv M_1 = M_2 = M_3$.

The standard model beta function is modified as follows\footnote{Note that the second term in the beta function of $\lambda$ is in disagreement with that of \cite{Gogoladze:2008ak} by a factor of 2.}  \cite{Chakrabortty:2008zh}
\begin{align}\label{eq:rgeLambdaTriplet}
\Delta \beta_\lambda &= \frac{1}{(4\pi)^2}\Big(12\lambda \text{Tr}\big[\mathbb{Y}_\Sigma^\dag \mathbb{Y}_\Sigma\big] - 5 \text{Tr}\big[\mathbb{Y}_\Sigma^\dag \mathbb{Y}_\Sigma\mathbb{Y}_\Sigma^\dag \mathbb{Y}_\Sigma\big]\Big)\\
\Delta \beta_{y_t} &= \frac{1}{(4\pi)^2} 3y_t \text{Tr}\big[\mathbb{Y}_\Sigma^\dag \mathbb{Y}_\Sigma\big] \\
\Delta \beta_{g} &= \frac{1}{(4\pi)^2} 4g^3\,.
\end{align}
The running of the Yukawa couplings are governed by the matrix-RGE:
\begin{multline}
[\beta_{\mathbb{Y}_\Sigma}]_{ij} = \frac{1}{(4\pi)^2}\Big[\frac{5}{2} (\mathbb{Y}_\Sigma \mathbb{Y}_\Sigma^\dag \mathbb{Y}_\Sigma)_{ij} \\ + (\mathbb{Y}_\Sigma)_{ij}\Big(3\text{Tr}\big[\mathbb{Y}_\Sigma^\dag \mathbb{Y}_\Sigma\big]
 + 3y_t^2 - \frac{3}{4}g'^2-\frac{33}{4}g^2\Big)\Big]\,,
\end{multline}
from which it follows the beta function of the trace
\begin{multline}\label{eq:rgeTrYYtriplet}
\beta_{\text{Tr}[\mathbb{Y}^\dag_\Sigma \mathbb{Y}_\Sigma]} = \frac{1}{(4\pi)^2}\Big[5\text{Tr}[\mathbb{Y}_\Sigma^\dag \mathbb{Y}_\Sigma \mathbb{Y}_\Sigma^\dag \mathbb{Y}_\Sigma] 
+ 6\big(\text{Tr}\big[\mathbb{Y}_\Sigma^\dag \mathbb{Y}_\Sigma\big]\big)^2
 \\
 +\text{Tr}\big[\mathbb{Y}_\Sigma^\dag \mathbb{Y}_\Sigma\big]\big( 6y_t^2 - \frac{3}{2}g'^2-\frac{33}{2}g^2\big)\Big]\,.
\end{multline}
As in the previous section, we assume that one triplet couples dominantly to the Higgs giving the relation in (\ref{eq:domCouplRelation}). And, in the case of low scale type-III seesaw with large triplet Yukawa couplings we use the relation in (\ref{eq:TrYrelationship}).

\begin{figure}
\includegraphics[width=0.8\columnwidth]{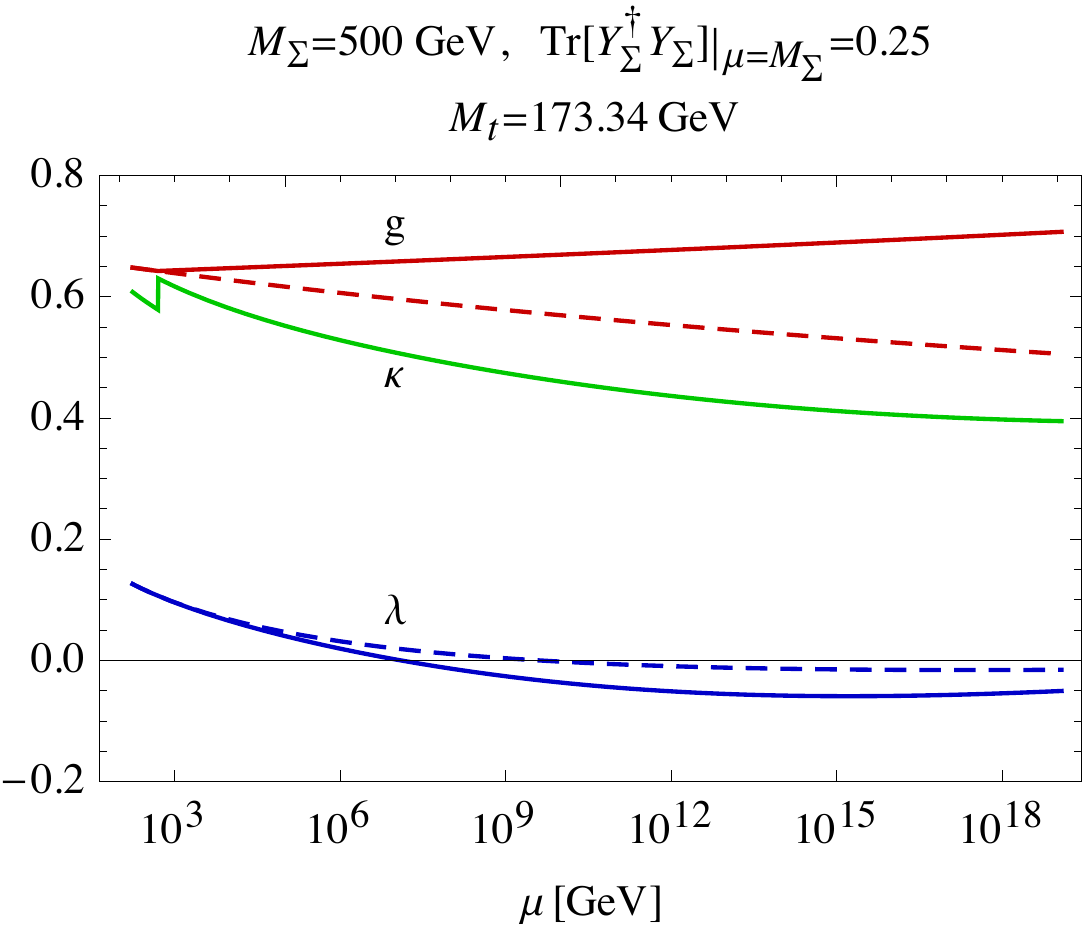}
\caption{The renormalization group evolution of the Higgs quartic coupling $\lambda$ (blue) and SU(2) gauge coupling $g$ (red) in the presence of isospin triplets (solid) and in the standard model (dashed).  In green is the thermal coefficient $\kappa$  defined in (\ref{eq:thermCoeffSM}) and modified by (\ref{eq:thermalCoefficientTriplet}).}\label{fig:rgetriplet}
\end{figure}

\begin{figure}
\includegraphics[width=0.8\columnwidth]{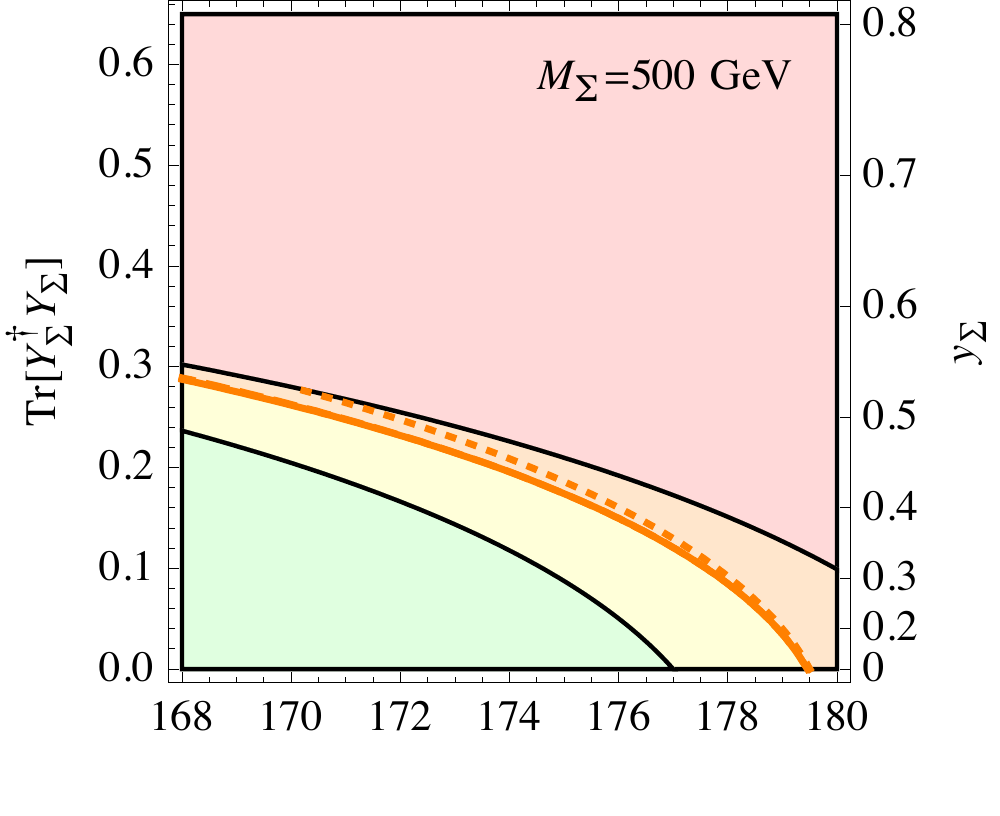}
\includegraphics[width=0.8\columnwidth]{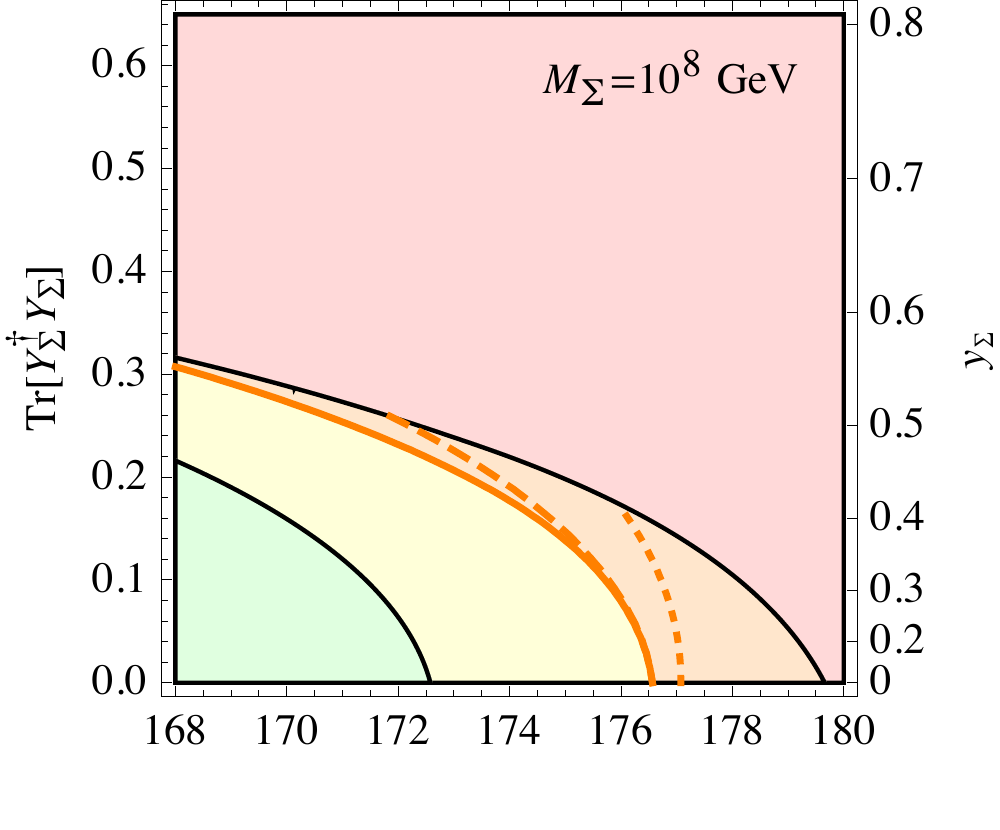}
\includegraphics[width=0.8\columnwidth]{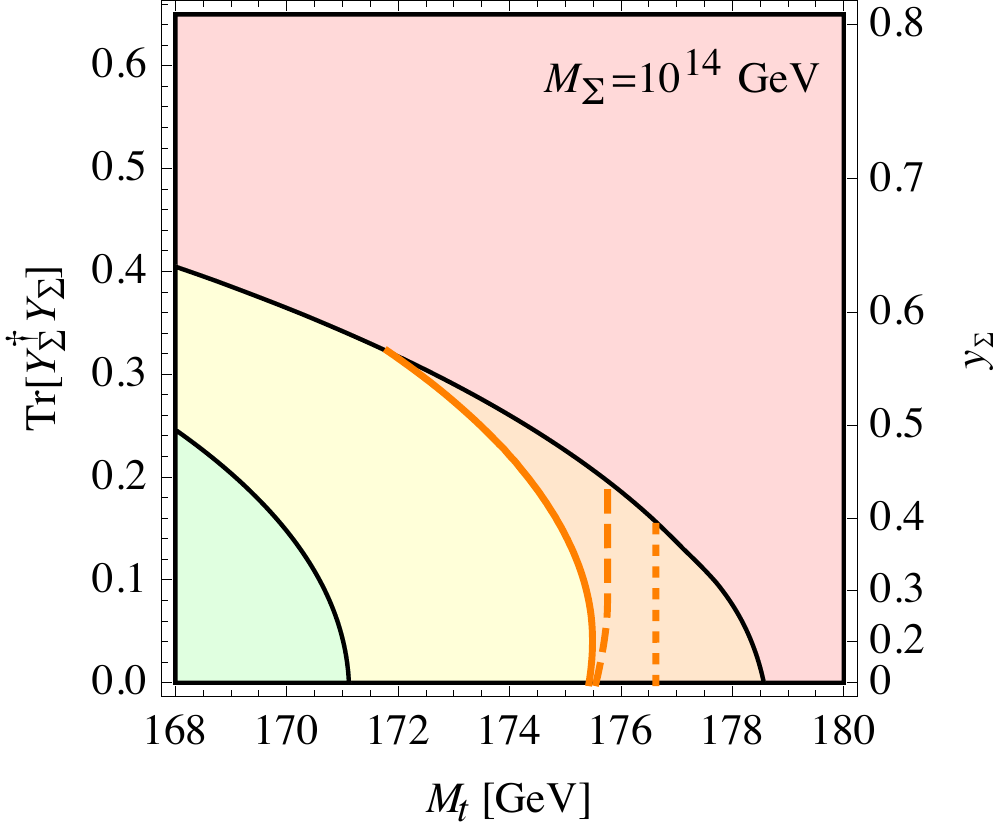}
\caption{Phase diagram of the standard model coupled to three fermionic isotriplets in the plane of $M_t$ and $y_\Sigma$, for representative values of $M_\Sigma$.  Coloring is equivalent to that in Fig. \ref{fig:PDsinglet}. Note that at $2\sigma$ C.L., the top quark pole mass is $\sim 171$--$175$ GeV.}\label{fig:phasespacetriplet}
\end{figure}

At temperatures above $M_\Sigma$, the Higgs thermal mass coefficient (\ref{eq:thermCoeffSM}) picks up an additional contribution from the fermionic isotriplets given by
\begin{equation}\label{eq:thermalCoefficientTriplet}
\delta \kappa^2 = \frac{1}{4}\text{Tr}\big[\mathbb{Y}_\nu^\dag \mathbb{Y}_\nu\big]\,.
\end{equation}

In Fig. \ref{fig:rgetriplet}, we compare the renormalization group running of $\lambda$ and SU(2) gauge coupling $g$ in this model with that in the standard model.  The nontrivial SU(2) charge of the triplet causes $g$ to run larger at high energies, driving the running of $\lambda$ in the positive direction.  The fermion triplet Yukawa coupling makes $\lambda$ run in the negative direction.  Therefore, there is a competition between the increased stabilizing effect of $g$ and the destabilizing effect of $y_\Sigma$.  

The phase diagram in the \mbox{$M_t$--$y_\Sigma$} plane for this model is shown in Fig. \ref{fig:phasespacetriplet}.  In addition to our findings in the case of right handed neutrinos, we make two additional observations for this case:

\begin{enumerate}
\setcounter{enumi}{3}
\item For small Yukawa couplings, the only effect of fermionic triplets is in the effective value of $g$ at high scales, indirectly affecting $\lambda$ to make the electroweak vacuum more stable.  This is evident by the enlarged region of absolute stability near small values of $y_\Sigma$.

\item The thermal stability bound is insensitive to $T_\text{max}$ for small $M_\Sigma$.  The reason is that the thermal crossing rate is controlled by the critical bubble energy in equation (\ref{eq:bubbenergy}).  Therefore, the highest rate occurs at a temperature $T_*$ when $|\lambda(\mu=T_*)|$ is largest.  According to the Fig. \ref{fig:rgetriplet}, $|\lambda|$ is maximized at around $T_* = 10^{15}$.  As a result, having $T_\text{max}$ larger than this $T_*$ does not change the bound.
\end{enumerate}

\section{Connection to observables}
In this section, we discuss the implications of our findings on experiments that can probe models of neutrino mass generation.  Sizable neutrino Yukawa couplings can allow the production of heavy neutrinos at colliders, leading to same sign dilepton final states.  The absence of these events in the CMS \cite{Khachatryan:2015gha,CMS:2015sea} and ATLAS \cite{Aad:2015xaa} detectors constrain the mixing of right handed neutrinos with the electron and muon neutrinos $|V_{lN}|^2 \sim \big((\mathbb{Y}_\nu)_{lN} v/M_R\big)^2$ to be in the range $10^{-3}$--$10^0$ for values of $M_R$ in the range $100$--$500$ GeV. The mixing with tau neutrinos is not constrained by LHC searches. For a recent analysis of bounds on heavy neutrino production at the LHC, see \cite{Das:2015toa, Deppisch:2015qwa}. Searches by experiments at the LHC \cite{Aad:2015cxa,CMS:2015mza} for fermion isotriplets via Drell-Yan production observe no evidence for their existence for masses up to $300$--$500$ GeV depending on their decay channels.

Electroweak precision data constrain $|V_{lN}|$ in the range $0.06$--$0.08$ \cite{delAguila:2008pw, Antusch:2014woa}. Broadly speaking, this leads to a constraint on the Yukawa coupling of $|(\mathbb{Y}_\nu)_{lN}| \lesssim 0.4 (M_R/\text{TeV})$ with $l=\{e,\mu,\tau\}$ and $N=\{1,2,3\}$. For a recent analysis of heavy neutrino effects on electroweak precision observables, lepton number and flavour violating decays, see \cite{Akhmedov:2013hec}.

Hierarchical thermal leptogenesis puts a bound on the mass of the lightest right handed neutrino and on the reheating temperature $T_{RH} \gtrsim 2 \times 10^{9}$ GeV \cite{Davidson:2002qv,Giudice:2003jh}. Our analysis shows that such high temperatures are allowed in the early Universe by stability bounds.

\section{Conclusions}
In this paper, we derived bounds by considering absolute, meta-, and early Universe thermal stability of the electroweak vacuum in models containing either right handed neutrinos or fermionic isotriplets.  To parametrize the effect of large neutrino Yukawa couplings in the renormalization group evolution equations for low scale type-I and III seesaw models by a single parameter, we demonstrate the validity of the approximate relationship (\ref{eq:TrYrelationship}) between traces of Yukawa coupling matrices.  Within the uncertainty of the measured value of the top quark mass, there is significant variation in the stability bounds on the neutrino Yukawa couplings.  We confirm the findings by \cite{Gogoladze:2008ak,He:2012ub} that for small isotriplet Yukawa couplings, its modification of the beta function of $g$ helps to stabilize the electroweak vacuum.

\acknowledgements
We thank Xun-Jie Xu for extensive discussions about flavor symmetries.  BR acknowledges the support from the Alexander von Humboldt Foundation.
\newpage
\bibliography{nuVacStab}

\begin{thebibliography}{48}%
\makeatletter
\providecommand \@ifxundefined [1]{%
 \@ifx{#1\undefined}
}%
\providecommand \@ifnum [1]{%
 \ifnum #1\expandafter \@firstoftwo
 \else \expandafter \@secondoftwo
 \fi
}%
\providecommand \@ifx [1]{%
 \ifx #1\expandafter \@firstoftwo
 \else \expandafter \@secondoftwo
 \fi
}%
\providecommand \natexlab [1]{#1}%
\providecommand \enquote  [1]{``#1''}%
\providecommand \bibnamefont  [1]{#1}%
\providecommand \bibfnamefont [1]{#1}%
\providecommand \citenamefont [1]{#1}%
\providecommand \href@noop [0]{\@secondoftwo}%
\providecommand \href [0]{\begingroup \@sanitize@url \@href}%
\providecommand \@href[1]{\@@startlink{#1}\@@href}%
\providecommand \@@href[1]{\endgroup#1\@@endlink}%
\providecommand \@sanitize@url [0]{\catcode `\\12\catcode `\$12\catcode
  `\&12\catcode `\#12\catcode `\^12\catcode `\_12\catcode `\%12\relax}%
\providecommand \@@startlink[1]{}%
\providecommand \@@endlink[0]{}%
\providecommand \url  [0]{\begingroup\@sanitize@url \@url }%
\providecommand \@url [1]{\endgroup\@href {#1}{\urlprefix }}%
\providecommand \urlprefix  [0]{URL }%
\providecommand \Eprint [0]{\href }%
\providecommand \doibase [0]{http://dx.doi.org/}%
\providecommand \selectlanguage [0]{\@gobble}%
\providecommand \bibinfo  [0]{\@secondoftwo}%
\providecommand \bibfield  [0]{\@secondoftwo}%
\providecommand \translation [1]{[#1]}%
\providecommand \BibitemOpen [0]{}%
\providecommand \bibitemStop [0]{}%
\providecommand \bibitemNoStop [0]{.\EOS\space}%
\providecommand \EOS [0]{\spacefactor3000\relax}%
\providecommand \BibitemShut  [1]{\csname bibitem#1\endcsname}%
\let\auto@bib@innerbib\@empty
\bibitem [{\citenamefont {Aad}\ \emph {et~al.}(2013)\citenamefont {Aad} \emph
  {et~al.}}]{Aad:2012tfa}%
  \BibitemOpen
  \bibfield  {author} {\bibinfo {author} {\bibfnamefont {G.}~\bibnamefont
  {Aad}} \emph {et~al.} (\bibinfo {collaboration} {ATLAS}),\ }\href {\doibase
  10.1016/j.physletb.2012.08.020} {\bibfield  {journal} {\bibinfo  {journal}
  {Phys. Lett.}\ }\textbf {\bibinfo {volume} {B716}},\ \bibinfo {pages} {1}
  (\bibinfo {year} {2013})},\ \Eprint {http://arxiv.org/abs/1207.7214}
  {arXiv:1207.7214 [hep-ex]} \BibitemShut {NoStop}%
\bibitem [{\citenamefont {Chatrchyan}\ \emph {et~al.}(2013)\citenamefont
  {Chatrchyan} \emph {et~al.}}]{Chatrchyan:2012xdj}%
  \BibitemOpen
  \bibfield  {author} {\bibinfo {author} {\bibfnamefont {S.}~\bibnamefont
  {Chatrchyan}} \emph {et~al.} (\bibinfo {collaboration} {CMS}),\ }\href
  {\doibase 10.1016/j.physletb.2012.08.021} {\bibfield  {journal} {\bibinfo
  {journal} {Phys. Lett.}\ }\textbf {\bibinfo {volume} {B716}},\ \bibinfo
  {pages} {30} (\bibinfo {year} {2013})},\ \Eprint
  {http://arxiv.org/abs/1207.7235} {arXiv:1207.7235 [hep-ex]} \BibitemShut
  {NoStop}%
\bibitem [{\citenamefont {Lindner}\ \emph {et~al.}(1989)\citenamefont
  {Lindner}, \citenamefont {Sher},\ and\ \citenamefont
  {Zaglauer}}]{Lindner:1988ww}%
  \BibitemOpen
  \bibfield  {author} {\bibinfo {author} {\bibfnamefont {M.}~\bibnamefont
  {Lindner}}, \bibinfo {author} {\bibfnamefont {M.}~\bibnamefont {Sher}}, \
  and\ \bibinfo {author} {\bibfnamefont {H.~W.}\ \bibnamefont {Zaglauer}},\
  }\href {\doibase 10.1016/0370-2693(89)90540-6} {\bibfield  {journal}
  {\bibinfo  {journal} {Phys. Lett.}\ }\textbf {\bibinfo {volume} {B228}},\
  \bibinfo {pages} {139} (\bibinfo {year} {1989})}\BibitemShut {NoStop}%
\bibitem [{\citenamefont {Buttazzo}\ \emph {et~al.}(2013)\citenamefont
  {Buttazzo}, \citenamefont {Degrassi}, \citenamefont {Giardino}, \citenamefont
  {Giudice}, \citenamefont {Sala}, \citenamefont {Salvio},\ and\ \citenamefont
  {Strumia}}]{Buttazzo:2013uya}%
  \BibitemOpen
  \bibfield  {author} {\bibinfo {author} {\bibfnamefont {D.}~\bibnamefont
  {Buttazzo}}, \bibinfo {author} {\bibfnamefont {G.}~\bibnamefont {Degrassi}},
  \bibinfo {author} {\bibfnamefont {P.~P.}\ \bibnamefont {Giardino}}, \bibinfo
  {author} {\bibfnamefont {G.~F.}\ \bibnamefont {Giudice}}, \bibinfo {author}
  {\bibfnamefont {F.}~\bibnamefont {Sala}}, \bibinfo {author} {\bibfnamefont
  {A.}~\bibnamefont {Salvio}}, \ and\ \bibinfo {author} {\bibfnamefont
  {A.}~\bibnamefont {Strumia}},\ }\href {\doibase 10.1007/JHEP12(2013)089}
  {\bibfield  {journal} {\bibinfo  {journal} {JHEP}\ }\textbf {\bibinfo
  {volume} {12}},\ \bibinfo {pages} {089} (\bibinfo {year} {2013})},\ \Eprint
  {http://arxiv.org/abs/1307.3536} {arXiv:1307.3536 [hep-ph]} \BibitemShut
  {NoStop}%
\bibitem [{\citenamefont {Anderson}(1990)}]{Anderson:1990aa}%
  \BibitemOpen
  \bibfield  {author} {\bibinfo {author} {\bibfnamefont {G.~W.}\ \bibnamefont
  {Anderson}},\ }\href {\doibase 10.1016/0370-2693(90)90849-2} {\bibfield
  {journal} {\bibinfo  {journal} {Phys. Lett.}\ }\textbf {\bibinfo {volume}
  {B243}},\ \bibinfo {pages} {265} (\bibinfo {year} {1990})}\BibitemShut
  {NoStop}%
\bibitem [{\citenamefont {Arnold}\ and\ \citenamefont
  {Vokos}(1991)}]{Arnold:1991cv}%
  \BibitemOpen
  \bibfield  {author} {\bibinfo {author} {\bibfnamefont {P.~B.}\ \bibnamefont
  {Arnold}}\ and\ \bibinfo {author} {\bibfnamefont {S.}~\bibnamefont {Vokos}},\
  }\href {\doibase 10.1103/PhysRevD.44.3620} {\bibfield  {journal} {\bibinfo
  {journal} {Phys. Rev.}\ }\textbf {\bibinfo {volume} {D44}},\ \bibinfo {pages}
  {3620} (\bibinfo {year} {1991})}\BibitemShut {NoStop}%
\bibitem [{\citenamefont {Espinosa}\ and\ \citenamefont
  {Quiros}(1995)}]{Espinosa:1995se}%
  \BibitemOpen
  \bibfield  {author} {\bibinfo {author} {\bibfnamefont {J.~R.}\ \bibnamefont
  {Espinosa}}\ and\ \bibinfo {author} {\bibfnamefont {M.}~\bibnamefont
  {Quiros}},\ }\href {\doibase 10.1016/0370-2693(95)00572-3} {\bibfield
  {journal} {\bibinfo  {journal} {Phys. Lett.}\ }\textbf {\bibinfo {volume}
  {B353}},\ \bibinfo {pages} {257} (\bibinfo {year} {1995})},\ \Eprint
  {http://arxiv.org/abs/hep-ph/9504241} {arXiv:hep-ph/9504241 [hep-ph]}
  \BibitemShut {NoStop}%
\bibitem [{\citenamefont {Espinosa}\ \emph {et~al.}(2008)\citenamefont
  {Espinosa}, \citenamefont {Giudice},\ and\ \citenamefont
  {Riotto}}]{Espinosa:2007qp}%
  \BibitemOpen
  \bibfield  {author} {\bibinfo {author} {\bibfnamefont {J.~R.}\ \bibnamefont
  {Espinosa}}, \bibinfo {author} {\bibfnamefont {G.~F.}\ \bibnamefont
  {Giudice}}, \ and\ \bibinfo {author} {\bibfnamefont {A.}~\bibnamefont
  {Riotto}},\ }\href {\doibase 10.1088/1475-7516/2008/05/002} {\bibfield
  {journal} {\bibinfo  {journal} {JCAP}\ }\textbf {\bibinfo {volume} {0805}},\
  \bibinfo {pages} {002} (\bibinfo {year} {2008})},\ \Eprint
  {http://arxiv.org/abs/0710.2484} {arXiv:0710.2484 [hep-ph]} \BibitemShut
  {NoStop}%
\bibitem [{\citenamefont {Delle~Rose}\ \emph
  {et~al.}(2015{\natexlab{a}})\citenamefont {Delle~Rose}, \citenamefont
  {Marzo},\ and\ \citenamefont {Urbano}}]{Rose:2015lna}%
  \BibitemOpen
  \bibfield  {author} {\bibinfo {author} {\bibfnamefont {L.}~\bibnamefont
  {Delle~Rose}}, \bibinfo {author} {\bibfnamefont {C.}~\bibnamefont {Marzo}}, \
  and\ \bibinfo {author} {\bibfnamefont {A.}~\bibnamefont {Urbano}},\
  }\href@noop {} {\  (\bibinfo {year} {2015}{\natexlab{a}})},\ \Eprint
  {http://arxiv.org/abs/1507.06912} {arXiv:1507.06912 [hep-ph]} \BibitemShut
  {NoStop}%
\bibitem [{\citenamefont {Casas}\ \emph {et~al.}(2000)\citenamefont {Casas},
  \citenamefont {Di~Clemente}, \citenamefont {Ibarra},\ and\ \citenamefont
  {Quiros}}]{Casas:1999cd}%
  \BibitemOpen
  \bibfield  {author} {\bibinfo {author} {\bibfnamefont {J.~A.}\ \bibnamefont
  {Casas}}, \bibinfo {author} {\bibfnamefont {V.}~\bibnamefont {Di~Clemente}},
  \bibinfo {author} {\bibfnamefont {A.}~\bibnamefont {Ibarra}}, \ and\ \bibinfo
  {author} {\bibfnamefont {M.}~\bibnamefont {Quiros}},\ }\href {\doibase
  10.1103/PhysRevD.62.053005} {\bibfield  {journal} {\bibinfo  {journal} {Phys.
  Rev.}\ }\textbf {\bibinfo {volume} {D62}},\ \bibinfo {pages} {053005}
  (\bibinfo {year} {2000})},\ \Eprint {http://arxiv.org/abs/hep-ph/9904295}
  {arXiv:hep-ph/9904295 [hep-ph]} \BibitemShut {NoStop}%
\bibitem [{\citenamefont {Elias-Miro}\ \emph {et~al.}(2012)\citenamefont
  {Elias-Miro}, \citenamefont {Espinosa}, \citenamefont {Giudice},
  \citenamefont {Isidori}, \citenamefont {Riotto},\ and\ \citenamefont
  {Strumia}}]{EliasMiro:2011aa}%
  \BibitemOpen
  \bibfield  {author} {\bibinfo {author} {\bibfnamefont {J.}~\bibnamefont
  {Elias-Miro}}, \bibinfo {author} {\bibfnamefont {J.~R.}\ \bibnamefont
  {Espinosa}}, \bibinfo {author} {\bibfnamefont {G.~F.}\ \bibnamefont
  {Giudice}}, \bibinfo {author} {\bibfnamefont {G.}~\bibnamefont {Isidori}},
  \bibinfo {author} {\bibfnamefont {A.}~\bibnamefont {Riotto}}, \ and\ \bibinfo
  {author} {\bibfnamefont {A.}~\bibnamefont {Strumia}},\ }\href {\doibase
  10.1016/j.physletb.2012.02.013} {\bibfield  {journal} {\bibinfo  {journal}
  {Phys. Lett.}\ }\textbf {\bibinfo {volume} {B709}},\ \bibinfo {pages} {222}
  (\bibinfo {year} {2012})},\ \Eprint {http://arxiv.org/abs/1112.3022}
  {arXiv:1112.3022 [hep-ph]} \BibitemShut {NoStop}%
\bibitem [{\citenamefont {Rodejohann}\ and\ \citenamefont
  {Zhang}(2012)}]{Rodejohann:2012px}%
  \BibitemOpen
  \bibfield  {author} {\bibinfo {author} {\bibfnamefont {W.}~\bibnamefont
  {Rodejohann}}\ and\ \bibinfo {author} {\bibfnamefont {H.}~\bibnamefont
  {Zhang}},\ }\href {\doibase 10.1007/JHEP06(2012)022} {\bibfield  {journal}
  {\bibinfo  {journal} {JHEP}\ }\textbf {\bibinfo {volume} {06}},\ \bibinfo
  {pages} {022} (\bibinfo {year} {2012})},\ \Eprint
  {http://arxiv.org/abs/1203.3825} {arXiv:1203.3825 [hep-ph]} \BibitemShut
  {NoStop}%
\bibitem [{\citenamefont {Bambhaniya}\ \emph {et~al.}(2015)\citenamefont
  {Bambhaniya}, \citenamefont {Khan}, \citenamefont {Konar},\ and\
  \citenamefont {Mondal}}]{Bambhaniya:2014hla}%
  \BibitemOpen
  \bibfield  {author} {\bibinfo {author} {\bibfnamefont {G.}~\bibnamefont
  {Bambhaniya}}, \bibinfo {author} {\bibfnamefont {S.}~\bibnamefont {Khan}},
  \bibinfo {author} {\bibfnamefont {P.}~\bibnamefont {Konar}}, \ and\ \bibinfo
  {author} {\bibfnamefont {T.}~\bibnamefont {Mondal}},\ }\href {\doibase
  10.1103/PhysRevD.91.095007} {\bibfield  {journal} {\bibinfo  {journal} {Phys.
  Rev.}\ }\textbf {\bibinfo {volume} {D91}},\ \bibinfo {pages} {095007}
  (\bibinfo {year} {2015})},\ \Eprint {http://arxiv.org/abs/1411.6866}
  {arXiv:1411.6866 [hep-ph]} \BibitemShut {NoStop}%
\bibitem [{\citenamefont {Delle~Rose}\ \emph
  {et~al.}(2015{\natexlab{b}})\citenamefont {Delle~Rose}, \citenamefont
  {Marzo},\ and\ \citenamefont {Urbano}}]{Rose:2015fua}%
  \BibitemOpen
  \bibfield  {author} {\bibinfo {author} {\bibfnamefont {L.}~\bibnamefont
  {Delle~Rose}}, \bibinfo {author} {\bibfnamefont {C.}~\bibnamefont {Marzo}}, \
  and\ \bibinfo {author} {\bibfnamefont {A.}~\bibnamefont {Urbano}},\
  }\href@noop {} {\  (\bibinfo {year} {2015}{\natexlab{b}})},\ \Eprint
  {http://arxiv.org/abs/1506.03360} {arXiv:1506.03360 [hep-ph]} \BibitemShut
  {NoStop}%
\bibitem [{\citenamefont {Khan}\ \emph {et~al.}(2014)\citenamefont {Khan},
  \citenamefont {Goswami},\ and\ \citenamefont {Roy}}]{Khan:2012zw}%
  \BibitemOpen
  \bibfield  {author} {\bibinfo {author} {\bibfnamefont {S.}~\bibnamefont
  {Khan}}, \bibinfo {author} {\bibfnamefont {S.}~\bibnamefont {Goswami}}, \
  and\ \bibinfo {author} {\bibfnamefont {S.}~\bibnamefont {Roy}},\ }\href
  {\doibase 10.1103/PhysRevD.89.073021} {\bibfield  {journal} {\bibinfo
  {journal} {Phys. Rev.}\ }\textbf {\bibinfo {volume} {D89}},\ \bibinfo {pages}
  {073021} (\bibinfo {year} {2014})},\ \Eprint {http://arxiv.org/abs/1212.3694}
  {arXiv:1212.3694 [hep-ph]} \BibitemShut {NoStop}%
\bibitem [{\citenamefont {Gogoladze}\ \emph {et~al.}(2008)\citenamefont
  {Gogoladze}, \citenamefont {Okada},\ and\ \citenamefont
  {Shafi}}]{Gogoladze:2008ak}%
  \BibitemOpen
  \bibfield  {author} {\bibinfo {author} {\bibfnamefont {I.}~\bibnamefont
  {Gogoladze}}, \bibinfo {author} {\bibfnamefont {N.}~\bibnamefont {Okada}}, \
  and\ \bibinfo {author} {\bibfnamefont {Q.}~\bibnamefont {Shafi}},\ }\href
  {\doibase 10.1016/j.physletb.2008.08.023} {\bibfield  {journal} {\bibinfo
  {journal} {Phys. Lett.}\ }\textbf {\bibinfo {volume} {B668}},\ \bibinfo
  {pages} {121} (\bibinfo {year} {2008})},\ \Eprint
  {http://arxiv.org/abs/0805.2129} {arXiv:0805.2129 [hep-ph]} \BibitemShut
  {NoStop}%
\bibitem [{\citenamefont {He}\ \emph {et~al.}(2012)\citenamefont {He},
  \citenamefont {Okada},\ and\ \citenamefont {Shafi}}]{He:2012ub}%
  \BibitemOpen
  \bibfield  {author} {\bibinfo {author} {\bibfnamefont {B.}~\bibnamefont
  {He}}, \bibinfo {author} {\bibfnamefont {N.}~\bibnamefont {Okada}}, \ and\
  \bibinfo {author} {\bibfnamefont {Q.}~\bibnamefont {Shafi}},\ }\href
  {\doibase 10.1016/j.physletb.2012.08.012} {\bibfield  {journal} {\bibinfo
  {journal} {Phys. Lett.}\ }\textbf {\bibinfo {volume} {B716}},\ \bibinfo
  {pages} {197} (\bibinfo {year} {2012})},\ \Eprint
  {http://arxiv.org/abs/1205.4038} {arXiv:1205.4038 [hep-ph]} \BibitemShut
  {NoStop}%
\bibitem [{\citenamefont {Isidori}\ \emph {et~al.}(2001)\citenamefont
  {Isidori}, \citenamefont {Ridolfi},\ and\ \citenamefont
  {Strumia}}]{Isidori:2001bm}%
  \BibitemOpen
  \bibfield  {author} {\bibinfo {author} {\bibfnamefont {G.}~\bibnamefont
  {Isidori}}, \bibinfo {author} {\bibfnamefont {G.}~\bibnamefont {Ridolfi}}, \
  and\ \bibinfo {author} {\bibfnamefont {A.}~\bibnamefont {Strumia}},\ }\href
  {\doibase 10.1016/S0550-3213(01)00302-9} {\bibfield  {journal} {\bibinfo
  {journal} {Nucl. Phys.}\ }\textbf {\bibinfo {volume} {B609}},\ \bibinfo
  {pages} {387} (\bibinfo {year} {2001})},\ \Eprint
  {http://arxiv.org/abs/hep-ph/0104016} {arXiv:hep-ph/0104016 [hep-ph]}
  \BibitemShut {NoStop}%
\bibitem [{\citenamefont {Guth}\ and\ \citenamefont
  {Weinberg}(1981)}]{Guth:1981uk}%
  \BibitemOpen
  \bibfield  {author} {\bibinfo {author} {\bibfnamefont {A.~H.}\ \bibnamefont
  {Guth}}\ and\ \bibinfo {author} {\bibfnamefont {E.~J.}\ \bibnamefont
  {Weinberg}},\ }\href {\doibase 10.1103/PhysRevD.23.876} {\bibfield  {journal}
  {\bibinfo  {journal} {Phys. Rev.}\ }\textbf {\bibinfo {volume} {D23}},\
  \bibinfo {pages} {876} (\bibinfo {year} {1981})}\BibitemShut {NoStop}%
\bibitem [{\citenamefont {Kramers}(1940)}]{Kramers:1940zz}%
  \BibitemOpen
  \bibfield  {author} {\bibinfo {author} {\bibfnamefont {H.~A.}\ \bibnamefont
  {Kramers}},\ }\href {\doibase 10.1016/S0031-8914(40)90098-2} {\bibfield
  {journal} {\bibinfo  {journal} {Physica}\ }\textbf {\bibinfo {volume} {7}},\
  \bibinfo {pages} {284} (\bibinfo {year} {1940})}\BibitemShut {NoStop}%
\bibitem [{\citenamefont {Patel}\ and\ \citenamefont
  {Ramsey-Musolf}(2011)}]{Patel:2011th}%
  \BibitemOpen
  \bibfield  {author} {\bibinfo {author} {\bibfnamefont {H.~H.}\ \bibnamefont
  {Patel}}\ and\ \bibinfo {author} {\bibfnamefont {M.~J.}\ \bibnamefont
  {Ramsey-Musolf}},\ }\href {\doibase 10.1007/JHEP07(2011)029} {\bibfield
  {journal} {\bibinfo  {journal} {JHEP}\ }\textbf {\bibinfo {volume} {07}},\
  \bibinfo {pages} {029} (\bibinfo {year} {2011})},\ \Eprint
  {http://arxiv.org/abs/1101.4665} {arXiv:1101.4665 [hep-ph]} \BibitemShut
  {NoStop}%
\bibitem [{\citenamefont {Minkowski}(1977)}]{Minkowski:1977sc}%
  \BibitemOpen
  \bibfield  {author} {\bibinfo {author} {\bibfnamefont {P.}~\bibnamefont
  {Minkowski}},\ }\href {\doibase 10.1016/0370-2693(77)90435-X} {\bibfield
  {journal} {\bibinfo  {journal} {Phys. Lett.}\ }\textbf {\bibinfo {volume}
  {B67}},\ \bibinfo {pages} {421} (\bibinfo {year} {1977})}\BibitemShut
  {NoStop}%
\bibitem [{\citenamefont {Yanagida}(1979)}]{Yanagida:1979as}%
  \BibitemOpen
  \bibfield  {author} {\bibinfo {author} {\bibfnamefont {T.}~\bibnamefont
  {Yanagida}},\ }\bibfield  {booktitle} {\emph {\bibinfo {booktitle}
  {{Proceedings: Workshop on the Unified Theories and the Baryon Number in the
  Universe, Tsukuba, Japan, 13-14 Feb 1979}}},\ }\href@noop {} {\bibfield
  {journal} {\bibinfo  {journal} {Conf. Proc.}\ }\textbf {\bibinfo {volume}
  {C7902131}},\ \bibinfo {pages} {95} (\bibinfo {year} {1979})},\ \bibinfo
  {note} {[Conf. Proc.C7902131,95(1979)]}\BibitemShut {NoStop}%
\bibitem [{\citenamefont {Gell-Mann}\ \emph {et~al.}(1979)\citenamefont
  {Gell-Mann}, \citenamefont {Ramond},\ and\ \citenamefont
  {Slansky}}]{GellMann:1980vs}%
  \BibitemOpen
  \bibfield  {author} {\bibinfo {author} {\bibfnamefont {M.}~\bibnamefont
  {Gell-Mann}}, \bibinfo {author} {\bibfnamefont {P.}~\bibnamefont {Ramond}}, \
  and\ \bibinfo {author} {\bibfnamefont {R.}~\bibnamefont {Slansky}},\
  }\bibfield  {booktitle} {\emph {\bibinfo {booktitle} {{Supergravity Workshop
  Stony Brook, New York, September 27-28, 1979}}},\ }\href@noop {} {\bibfield
  {journal} {\bibinfo  {journal} {Conf. Proc.}\ }\textbf {\bibinfo {volume}
  {C790927}},\ \bibinfo {pages} {315} (\bibinfo {year} {1979})},\ \Eprint
  {http://arxiv.org/abs/1306.4669} {arXiv:1306.4669 [hep-th]} \BibitemShut
  {NoStop}%
\bibitem [{\citenamefont {Mohapatra}\ and\ \citenamefont
  {Senjanovic}(1980)}]{Mohapatra:1979ia}%
  \BibitemOpen
  \bibfield  {author} {\bibinfo {author} {\bibfnamefont {R.~N.}\ \bibnamefont
  {Mohapatra}}\ and\ \bibinfo {author} {\bibfnamefont {G.}~\bibnamefont
  {Senjanovic}},\ }\href {\doibase 10.1103/PhysRevLett.44.912} {\bibfield
  {journal} {\bibinfo  {journal} {Phys. Rev. Lett.}\ }\textbf {\bibinfo
  {volume} {44}},\ \bibinfo {pages} {912} (\bibinfo {year} {1980})}\BibitemShut
  {NoStop}%
\bibitem [{\citenamefont {Mohapatra}\ and\ \citenamefont
  {Valle}(1986)}]{Mohapatra:1986bd}%
  \BibitemOpen
  \bibfield  {author} {\bibinfo {author} {\bibfnamefont {R.~N.}\ \bibnamefont
  {Mohapatra}}\ and\ \bibinfo {author} {\bibfnamefont {J.~W.~F.}\ \bibnamefont
  {Valle}},\ }\bibfield  {booktitle} {\emph {\bibinfo {booktitle}
  {{Proceedings, 23RD International Conference on High Energy Physics, JULY
  16-23, 1986, Berkeley, CA}}},\ }\href {\doibase 10.1103/PhysRevD.34.1642}
  {\bibfield  {journal} {\bibinfo  {journal} {Phys. Rev.}\ }\textbf {\bibinfo
  {volume} {D34}},\ \bibinfo {pages} {1642} (\bibinfo {year}
  {1986})}\BibitemShut {NoStop}%
\bibitem [{\citenamefont {Gonzalez-Garcia}\ and\ \citenamefont
  {Valle}(1989)}]{GonzalezGarcia:1988rw}%
  \BibitemOpen
  \bibfield  {author} {\bibinfo {author} {\bibfnamefont {M.~C.}\ \bibnamefont
  {Gonzalez-Garcia}}\ and\ \bibinfo {author} {\bibfnamefont {J.~W.~F.}\
  \bibnamefont {Valle}},\ }\href {\doibase 10.1016/0370-2693(89)91131-3}
  {\bibfield  {journal} {\bibinfo  {journal} {Phys. Lett.}\ }\textbf {\bibinfo
  {volume} {B216}},\ \bibinfo {pages} {360} (\bibinfo {year}
  {1989})}\BibitemShut {NoStop}%
\bibitem [{\citenamefont {Malinsky}\ \emph {et~al.}(2005)\citenamefont
  {Malinsky}, \citenamefont {Romao},\ and\ \citenamefont
  {Valle}}]{Malinsky:2005bi}%
  \BibitemOpen
  \bibfield  {author} {\bibinfo {author} {\bibfnamefont {M.}~\bibnamefont
  {Malinsky}}, \bibinfo {author} {\bibfnamefont {J.~C.}\ \bibnamefont {Romao}},
  \ and\ \bibinfo {author} {\bibfnamefont {J.~W.~F.}\ \bibnamefont {Valle}},\
  }\href {\doibase 10.1103/PhysRevLett.95.161801} {\bibfield  {journal}
  {\bibinfo  {journal} {Phys. Rev. Lett.}\ }\textbf {\bibinfo {volume} {95}},\
  \bibinfo {pages} {161801} (\bibinfo {year} {2005})},\ \Eprint
  {http://arxiv.org/abs/hep-ph/0506296} {arXiv:hep-ph/0506296 [hep-ph]}
  \BibitemShut {NoStop}%
\bibitem [{\citenamefont {Grzadkowski}\ and\ \citenamefont
  {Lindner}(1987)}]{Grzadkowski:1987tf}%
  \BibitemOpen
  \bibfield  {author} {\bibinfo {author} {\bibfnamefont {B.}~\bibnamefont
  {Grzadkowski}}\ and\ \bibinfo {author} {\bibfnamefont {M.}~\bibnamefont
  {Lindner}},\ }\href {\doibase 10.1016/0370-2693(87)90458-8} {\bibfield
  {journal} {\bibinfo  {journal} {Phys. Lett.}\ }\textbf {\bibinfo {volume}
  {B193}},\ \bibinfo {pages} {71} (\bibinfo {year} {1987})}\BibitemShut
  {NoStop}%
\bibitem [{\citenamefont {Pirogov}\ and\ \citenamefont
  {Zenin}(1999)}]{Pirogov:1998tj}%
  \BibitemOpen
  \bibfield  {author} {\bibinfo {author} {\bibfnamefont {{\relax Yu}.~F.}\
  \bibnamefont {Pirogov}}\ and\ \bibinfo {author} {\bibfnamefont {O.~V.}\
  \bibnamefont {Zenin}},\ }\href {\doibase 10.1007/s100520050602} {\bibfield
  {journal} {\bibinfo  {journal} {Eur. Phys. J.}\ }\textbf {\bibinfo {volume}
  {C10}},\ \bibinfo {pages} {629} (\bibinfo {year} {1999})},\ \Eprint
  {http://arxiv.org/abs/hep-ph/9808396} {arXiv:hep-ph/9808396 [hep-ph]}
  \BibitemShut {NoStop}%
\bibitem [{\citenamefont {Casas}\ and\ \citenamefont
  {Ibarra}(2001)}]{Casas:2001sr}%
  \BibitemOpen
  \bibfield  {author} {\bibinfo {author} {\bibfnamefont {J.~A.}\ \bibnamefont
  {Casas}}\ and\ \bibinfo {author} {\bibfnamefont {A.}~\bibnamefont {Ibarra}},\
  }\href {\doibase 10.1016/S0550-3213(01)00475-8} {\bibfield  {journal}
  {\bibinfo  {journal} {Nucl. Phys.}\ }\textbf {\bibinfo {volume} {B618}},\
  \bibinfo {pages} {171} (\bibinfo {year} {2001})},\ \Eprint
  {http://arxiv.org/abs/hep-ph/0103065} {arXiv:hep-ph/0103065 [hep-ph]}
  \BibitemShut {NoStop}%
\bibitem [{\citenamefont {Pascoli}\ \emph {et~al.}(2003)\citenamefont
  {Pascoli}, \citenamefont {Petcov},\ and\ \citenamefont
  {Yaguna}}]{Pascoli:2003rq}%
  \BibitemOpen
  \bibfield  {author} {\bibinfo {author} {\bibfnamefont {S.}~\bibnamefont
  {Pascoli}}, \bibinfo {author} {\bibfnamefont {S.~T.}\ \bibnamefont {Petcov}},
  \ and\ \bibinfo {author} {\bibfnamefont {C.~E.}\ \bibnamefont {Yaguna}},\
  }\href {\doibase 10.1016/S0370-2693(03)00698-1} {\bibfield  {journal}
  {\bibinfo  {journal} {Phys. Lett.}\ }\textbf {\bibinfo {volume} {B564}},\
  \bibinfo {pages} {241} (\bibinfo {year} {2003})},\ \Eprint
  {http://arxiv.org/abs/hep-ph/0301095} {arXiv:hep-ph/0301095 [hep-ph]}
  \BibitemShut {NoStop}%
\bibitem [{\citenamefont {Aad}\ \emph {et~al.}(2014)\citenamefont {Aad} \emph
  {et~al.}}]{ATLAS:2014wva}%
  \BibitemOpen
  \bibfield  {author} {\bibinfo {author} {\bibfnamefont {G.}~\bibnamefont
  {Aad}} \emph {et~al.} (\bibinfo {collaboration} {ATLAS, CDF, CMS, D0
  Collaborations}),\ }\href@noop {} {\  (\bibinfo {year} {2014})},\ \Eprint
  {http://arxiv.org/abs/1403.4427} {arXiv:1403.4427 [hep-ex]} \BibitemShut
  {NoStop}%
\bibitem [{\citenamefont {Aad}\ \emph {et~al.}(2015{\natexlab{a}})\citenamefont
  {Aad} \emph {et~al.}}]{Aad:2015zhl}%
  \BibitemOpen
  \bibfield  {author} {\bibinfo {author} {\bibfnamefont {G.}~\bibnamefont
  {Aad}} \emph {et~al.} (\bibinfo {collaboration} {ATLAS, CMS}),\ }\href
  {\doibase 10.1103/PhysRevLett.114.191803} {\bibfield  {journal} {\bibinfo
  {journal} {Phys. Rev. Lett.}\ }\textbf {\bibinfo {volume} {114}},\ \bibinfo
  {pages} {191803} (\bibinfo {year} {2015}{\natexlab{a}})},\ \Eprint
  {http://arxiv.org/abs/1503.07589} {arXiv:1503.07589 [hep-ex]} \BibitemShut
  {NoStop}%
\bibitem [{\citenamefont {Foot}\ \emph {et~al.}(1989)\citenamefont {Foot},
  \citenamefont {Lew}, \citenamefont {He},\ and\ \citenamefont
  {Joshi}}]{Foot:1988aq}%
  \BibitemOpen
  \bibfield  {author} {\bibinfo {author} {\bibfnamefont {R.}~\bibnamefont
  {Foot}}, \bibinfo {author} {\bibfnamefont {H.}~\bibnamefont {Lew}}, \bibinfo
  {author} {\bibfnamefont {X.~G.}\ \bibnamefont {He}}, \ and\ \bibinfo {author}
  {\bibfnamefont {G.~C.}\ \bibnamefont {Joshi}},\ }\href {\doibase
  10.1007/BF01415558} {\bibfield  {journal} {\bibinfo  {journal} {Z. Phys.}\
  }\textbf {\bibinfo {volume} {C44}},\ \bibinfo {pages} {441} (\bibinfo {year}
  {1989})}\BibitemShut {NoStop}%
\bibitem [{\citenamefont {Chakrabortty}\ \emph {et~al.}(2009)\citenamefont
  {Chakrabortty}, \citenamefont {Dighe}, \citenamefont {Goswami},\ and\
  \citenamefont {Ray}}]{Chakrabortty:2008zh}%
  \BibitemOpen
  \bibfield  {author} {\bibinfo {author} {\bibfnamefont {J.}~\bibnamefont
  {Chakrabortty}}, \bibinfo {author} {\bibfnamefont {A.}~\bibnamefont {Dighe}},
  \bibinfo {author} {\bibfnamefont {S.}~\bibnamefont {Goswami}}, \ and\
  \bibinfo {author} {\bibfnamefont {S.}~\bibnamefont {Ray}},\ }\href {\doibase
  10.1016/j.nuclphysb.2009.05.016} {\bibfield  {journal} {\bibinfo  {journal}
  {Nucl. Phys.}\ }\textbf {\bibinfo {volume} {B820}},\ \bibinfo {pages} {116}
  (\bibinfo {year} {2009})},\ \Eprint {http://arxiv.org/abs/0812.2776}
  {arXiv:0812.2776 [hep-ph]} \BibitemShut {NoStop}%
\bibitem [{\citenamefont {Khachatryan}\ \emph
  {et~al.}(2015{\natexlab{a}})\citenamefont {Khachatryan} \emph
  {et~al.}}]{Khachatryan:2015gha}%
  \BibitemOpen
  \bibfield  {author} {\bibinfo {author} {\bibfnamefont {V.}~\bibnamefont
  {Khachatryan}} \emph {et~al.} (\bibinfo {collaboration} {CMS}),\ }\href
  {\doibase 10.1016/j.physletb.2015.06.070} {\bibfield  {journal} {\bibinfo
  {journal} {Phys. Lett.}\ }\textbf {\bibinfo {volume} {B748}},\ \bibinfo
  {pages} {144} (\bibinfo {year} {2015}{\natexlab{a}})},\ \Eprint
  {http://arxiv.org/abs/1501.05566} {arXiv:1501.05566 [hep-ex]} \BibitemShut
  {NoStop}%
\bibitem [{\citenamefont {Khachatryan}\ \emph
  {et~al.}(2015{\natexlab{b}})\citenamefont {Khachatryan} \emph
  {et~al.}}]{CMS:2015sea}%
  \BibitemOpen
  \bibfield  {author} {\bibinfo {author} {\bibfnamefont {V.}~\bibnamefont
  {Khachatryan}} \emph {et~al.} (\bibinfo {collaboration} {CMS}),\ }\href@noop
  {} {\  (\bibinfo {year} {2015}{\natexlab{b}})}\BibitemShut {NoStop}%
\bibitem [{\citenamefont {Aad}\ \emph {et~al.}(2015{\natexlab{b}})\citenamefont
  {Aad} \emph {et~al.}}]{Aad:2015xaa}%
  \BibitemOpen
  \bibfield  {author} {\bibinfo {author} {\bibfnamefont {G.}~\bibnamefont
  {Aad}} \emph {et~al.} (\bibinfo {collaboration} {ATLAS}),\ }\href {\doibase
  10.1007/JHEP07(2015)162} {\bibfield  {journal} {\bibinfo  {journal} {JHEP}\
  }\textbf {\bibinfo {volume} {07}},\ \bibinfo {pages} {162} (\bibinfo {year}
  {2015}{\natexlab{b}})},\ \Eprint {http://arxiv.org/abs/1506.06020}
  {arXiv:1506.06020 [hep-ex]} \BibitemShut {NoStop}%
\bibitem [{\citenamefont {Das}\ and\ \citenamefont
  {Okada}(2015)}]{Das:2015toa}%
  \BibitemOpen
  \bibfield  {author} {\bibinfo {author} {\bibfnamefont {A.}~\bibnamefont
  {Das}}\ and\ \bibinfo {author} {\bibfnamefont {N.}~\bibnamefont {Okada}},\
  }\href@noop {} {\  (\bibinfo {year} {2015})},\ \Eprint
  {http://arxiv.org/abs/1510.04790} {arXiv:1510.04790 [hep-ph]} \BibitemShut
  {NoStop}%
\bibitem [{\citenamefont {Deppisch}\ \emph {et~al.}(2015)\citenamefont
  {Deppisch}, \citenamefont {Bhupal~Dev},\ and\ \citenamefont
  {Pilaftsis}}]{Deppisch:2015qwa}%
  \BibitemOpen
  \bibfield  {author} {\bibinfo {author} {\bibfnamefont {F.~F.}\ \bibnamefont
  {Deppisch}}, \bibinfo {author} {\bibfnamefont {P.~S.}\ \bibnamefont
  {Bhupal~Dev}}, \ and\ \bibinfo {author} {\bibfnamefont {A.}~\bibnamefont
  {Pilaftsis}},\ }\href {\doibase 10.1088/1367-2630/17/7/075019} {\bibfield
  {journal} {\bibinfo  {journal} {New J. Phys.}\ }\textbf {\bibinfo {volume}
  {17}},\ \bibinfo {pages} {075019} (\bibinfo {year} {2015})},\ \Eprint
  {http://arxiv.org/abs/1502.06541} {arXiv:1502.06541 [hep-ph]} \BibitemShut
  {NoStop}%
\bibitem [{\citenamefont {Aad}\ \emph {et~al.}(2015{\natexlab{c}})\citenamefont
  {Aad} \emph {et~al.}}]{Aad:2015cxa}%
  \BibitemOpen
  \bibfield  {author} {\bibinfo {author} {\bibfnamefont {G.}~\bibnamefont
  {Aad}} \emph {et~al.} (\bibinfo {collaboration} {ATLAS}),\ }\href {\doibase
  10.1103/PhysRevD.92.032001} {\bibfield  {journal} {\bibinfo  {journal} {Phys.
  Rev.}\ }\textbf {\bibinfo {volume} {D92}},\ \bibinfo {pages} {032001}
  (\bibinfo {year} {2015}{\natexlab{c}})},\ \Eprint
  {http://arxiv.org/abs/1506.01839} {arXiv:1506.01839 [hep-ex]} \BibitemShut
  {NoStop}%
\bibitem [{\citenamefont {Khachatryan}\ \emph
  {et~al.}(2015{\natexlab{c}})\citenamefont {Khachatryan} \emph
  {et~al.}}]{CMS:2015mza}%
  \BibitemOpen
  \bibfield  {author} {\bibinfo {author} {\bibfnamefont {V.}~\bibnamefont
  {Khachatryan}} \emph {et~al.} (\bibinfo {collaboration} {CMS}),\ }\href@noop
  {} {\  (\bibinfo {year} {2015}{\natexlab{c}})}\BibitemShut {NoStop}%
\bibitem [{\citenamefont {del Aguila}\ \emph {et~al.}(2008)\citenamefont {del
  Aguila}, \citenamefont {de~Blas},\ and\ \citenamefont
  {Perez-Victoria}}]{delAguila:2008pw}%
  \BibitemOpen
  \bibfield  {author} {\bibinfo {author} {\bibfnamefont {F.}~\bibnamefont {del
  Aguila}}, \bibinfo {author} {\bibfnamefont {J.}~\bibnamefont {de~Blas}}, \
  and\ \bibinfo {author} {\bibfnamefont {M.}~\bibnamefont {Perez-Victoria}},\
  }\href {\doibase 10.1103/PhysRevD.78.013010} {\bibfield  {journal} {\bibinfo
  {journal} {Phys. Rev.}\ }\textbf {\bibinfo {volume} {D78}},\ \bibinfo {pages}
  {013010} (\bibinfo {year} {2008})},\ \Eprint {http://arxiv.org/abs/0803.4008}
  {arXiv:0803.4008 [hep-ph]} \BibitemShut {NoStop}%
\bibitem [{\citenamefont {Antusch}\ and\ \citenamefont
  {Fischer}(2014)}]{Antusch:2014woa}%
  \BibitemOpen
  \bibfield  {author} {\bibinfo {author} {\bibfnamefont {S.}~\bibnamefont
  {Antusch}}\ and\ \bibinfo {author} {\bibfnamefont {O.}~\bibnamefont
  {Fischer}},\ }\href {\doibase 10.1007/JHEP10(2014)094} {\bibfield  {journal}
  {\bibinfo  {journal} {JHEP}\ }\textbf {\bibinfo {volume} {10}},\ \bibinfo
  {pages} {94} (\bibinfo {year} {2014})},\ \Eprint
  {http://arxiv.org/abs/1407.6607} {arXiv:1407.6607 [hep-ph]} \BibitemShut
  {NoStop}%
\bibitem [{\citenamefont {Akhmedov}\ \emph {et~al.}(2013)\citenamefont
  {Akhmedov}, \citenamefont {Kartavtsev}, \citenamefont {Lindner},
  \citenamefont {Michaels},\ and\ \citenamefont {Smirnov}}]{Akhmedov:2013hec}%
  \BibitemOpen
  \bibfield  {author} {\bibinfo {author} {\bibfnamefont {E.}~\bibnamefont
  {Akhmedov}}, \bibinfo {author} {\bibfnamefont {A.}~\bibnamefont
  {Kartavtsev}}, \bibinfo {author} {\bibfnamefont {M.}~\bibnamefont {Lindner}},
  \bibinfo {author} {\bibfnamefont {L.}~\bibnamefont {Michaels}}, \ and\
  \bibinfo {author} {\bibfnamefont {J.}~\bibnamefont {Smirnov}},\ }\href
  {\doibase 10.1007/JHEP05(2013)081} {\bibfield  {journal} {\bibinfo  {journal}
  {JHEP}\ }\textbf {\bibinfo {volume} {05}},\ \bibinfo {pages} {081} (\bibinfo
  {year} {2013})},\ \Eprint {http://arxiv.org/abs/1302.1872} {arXiv:1302.1872
  [hep-ph]} \BibitemShut {NoStop}%
\bibitem [{\citenamefont {Davidson}\ and\ \citenamefont
  {Ibarra}(2002)}]{Davidson:2002qv}%
  \BibitemOpen
  \bibfield  {author} {\bibinfo {author} {\bibfnamefont {S.}~\bibnamefont
  {Davidson}}\ and\ \bibinfo {author} {\bibfnamefont {A.}~\bibnamefont
  {Ibarra}},\ }\href {\doibase 10.1016/S0370-2693(02)01735-5} {\bibfield
  {journal} {\bibinfo  {journal} {Phys. Lett.}\ }\textbf {\bibinfo {volume}
  {B535}},\ \bibinfo {pages} {25} (\bibinfo {year} {2002})},\ \Eprint
  {http://arxiv.org/abs/hep-ph/0202239} {arXiv:hep-ph/0202239 [hep-ph]}
  \BibitemShut {NoStop}%
\bibitem [{\citenamefont {Giudice}\ \emph {et~al.}(2004)\citenamefont
  {Giudice}, \citenamefont {Notari}, \citenamefont {Raidal}, \citenamefont
  {Riotto},\ and\ \citenamefont {Strumia}}]{Giudice:2003jh}%
  \BibitemOpen
  \bibfield  {author} {\bibinfo {author} {\bibfnamefont {G.~F.}\ \bibnamefont
  {Giudice}}, \bibinfo {author} {\bibfnamefont {A.}~\bibnamefont {Notari}},
  \bibinfo {author} {\bibfnamefont {M.}~\bibnamefont {Raidal}}, \bibinfo
  {author} {\bibfnamefont {A.}~\bibnamefont {Riotto}}, \ and\ \bibinfo {author}
  {\bibfnamefont {A.}~\bibnamefont {Strumia}},\ }\href {\doibase
  10.1016/j.nuclphysb.2004.02.019} {\bibfield  {journal} {\bibinfo  {journal}
  {Nucl. Phys.}\ }\textbf {\bibinfo {volume} {B685}},\ \bibinfo {pages} {89}
  (\bibinfo {year} {2004})},\ \Eprint {http://arxiv.org/abs/hep-ph/0310123}
  {arXiv:hep-ph/0310123 [hep-ph]} \BibitemShut {NoStop}%
\end{thebibliography}%

\end{document}